\documentclass[a4paper,11pt]{article}
\pdfoutput=1 % if your are submitting a pdflatex (i.e. if you have
\usepackage{jheppub} % for details on the use of the package, please
\usepackage[T1]{fontenc} % if needed
\usepackage{bm}% bold math
\usepackage{url}
\DeclareMathAlphabet{\pazocal}{OMS}{zplm}{m}{n}
\usepackage{braket}
\usepackage{amsmath}
\usepackage{amssymb}
\usepackage{leftidx}
\usepackage[colorinlistoftodos]{todonotes}
\usepackage{bm}
\usepackage{slashed}
\usepackage{calrsfs}
\usepackage{cancel}
\usepackage{dirtytalk}
\usepackage{upgreek}
\usepackage{hyperref}
\title{\boldmath Unconventional conformal invariance of maximal depth partially massless fields on $dS_{4}$ and its relation to complex partially massless SUSY}

\author[a]{Vasileios A. Letsios}

\affiliation[a]{Department of Mathematics, King’s College London,\\  Strand, London WC2R 2LS, UK}

\emailAdd{vasileios.letsios@kcl.ac.uk}

\abstract{Deser and Waldron have shown that maximal depth partially massless theories of higher (integer) spin on four-dimensional de Sitter spacetime ($dS_{4}$) possess infinitesimal symmetries generated by the conformal Killing vectors of $dS_{4}$. However, it was later shown by Barnich, Bekaert, and Grigoriev that these theories are not invariant under the conformal algebra $so(4,2)$. To get some insight into these seemingly contradicting results we write down the full set of infinitesimal transformations of the fields generated by the fifteen conformal Killing vectors of $dS_{4}$. In particular, although the infinitesimal transformations generated by the ten dS Killing vectors are well-known (these correspond to the conventional Lie derivatives), the transformations generated by the five non-Killing conformal Killing vectors were absent from the literature, and we show that they have an `unconventional' form. In the spin-2 case (partially massless graviton), we show that the field equations and the action are invariant under the unconventional conformal transformations. For spin $s >2$, the invariance is demonstrated only at the level of the field equations. For all spins $s \geq 2$, we reproduce the result that the symmetry algebra does not close on the conformal algebra, $so(4,2)$. This is due to the appearance of new higher-derivative symmetry transformations in the commutator of two unconventional conformal transformations. Our results concerning the closure of the full symmetry algebra are inconclusive. Then we shift focus to the question of supersymmetry (SUSY) on $dS_{4}$ and our objective is twofold. First, we uncover a non-interacting supermultiplet that consists of a complex partially massless spin-2 field and a complex spin-3/2 field on $dS_{4}$. Second, we showcase the appearance of the unconventional conformal symmetries in the commutator of two SUSY transformations. Thus, this commutator closes on an algebra that is neither $so(4,1)$ nor $so(4,2)$, while its full structure is an open question. More open questions arising from our findings are also discussed. }

\begin{document} 
\maketitle
\flushbottom

\section{Introduction}
 Studying field theories in de Sitter (dS) spacetime is interesting not only because of its relevance to inflationary cosmology but also because our Universe is currently undergoing a phase of accelerated spatial expansion~\cite{Perlmutter_1999,SloanDigitalSky,PlanckCollab, Galante}, i.e. our Universe might end as de Sitter space.

 Four-dimensional dS spacetime ($dS_{4}$), is the maximally symmetric solution of the vacuum Einstein equations with positive cosmological constant $\Lambda$,
 \begin{equation}
    R_{\mu \nu}-\frac{1}{2} g_{\mu \nu} R + \Lambda  g_{\mu \nu}=0,
\end{equation}
where $g_{\mu \nu}$ is the $dS_{4}$ metric tensor, while $R_{\mu \nu}$ is the Ricci tensor and $R$ is the Ricci scalar. We will work with units in which $\Lambda = 3$. The $dS_{4}$ Riemann tensor is given by
\begin{align}\label{Riemann}
   R_{\mu \nu \rho \sigma} = g_{\mu \rho}\, g_{\nu \sigma} - g_{\nu \rho} \,g_{\mu \sigma}. 
\end{align}

Integer-spin-$s$ fields on $dS_{4}$ can be described by totally symmetric tensors of rank $s$, $h_{\mu_{1}...\mu_{s}}$\footnote{Mixed-symmetry tensor fields have been discussed in Ref.~\cite{Mixed_Symmetry_dS}.}. These satisfy the on-shell conditions~\cite{STSHS, Deser_Waldron_null_propagation}:
\begin{align} \label{EOM massive}
    &\Big( \nabla^{\alpha} \nabla_{\alpha}-m^{2} +  (s-2)(s+1)-s \Big)h_{\mu_{1}...\mu_{s}} = 0, \nonumber\\
    & \nabla^{\alpha} h_{\alpha \mu_{2}...\mu_{s}}=0, \hspace{4mm} g^{\alpha \beta} h_{\alpha \beta \mu_{3}...\mu_{s}},
\end{align}
where $m$ is the mass parameter. In this article, our main investigations concern classical integer-spin fields. Throughout most of the article, we assume that $s \geq 2$.

The field equations~(\ref{EOM massive}) enjoy invariance under the dS isometries. {At the classical level, infinitesimal dS invariance of the field equations can be studied in terms of the standard action of isometries through infinitesimal active diffeomorphisms. In particular}, the generators of the dS algebra, $so(4,1)$, are the ten Killing vectors $\xi^{\mu}$ of $dS_{4}$
\begin{align}\label{KV equation}
    \nabla_{\mu}   \xi_{\nu}+\nabla_{\nu}\xi_{\mu}=0.
\end{align}
The Killing vectors $\xi$ act on solutions $h_{\mu_{1}...\mu_{s}}$ in terms of the Lie derivative
\begin{align}\label{Lie derivative}
    \mathcal{L}_{\xi} h_{\mu_{1}...\mu_{s}} = \xi^{\rho}\nabla_{\rho}h_{\mu_{1}...\mu_{s}} + s \nabla_{(\mu_{1}} \xi^{\sigma} h_{\mu_{2}...\mu_{s}) \sigma}.
\end{align}
The Lie derivative~(\ref{Lie derivative}) maps solutions $h_{\mu_{1}...\mu_{s}}$ to other solutions (as $\mathcal{L}_{\xi}$ commutes with covariant derivatives), and thus, it is a symmetry of the field equations~(\ref{EOM massive}). Also, since $[\mathcal{L}_{\xi} , \mathcal{L}_{\xi '}] = \mathcal{L}_{[\xi , \xi']}$ for any two dS Killing vectors $\xi , \xi'$, the Lie derivatives generate a representation of $so(4,1)$ on the solution space of Eq.~(\ref{EOM massive}).

%%%%%%%%%%%%%%%%%%%%%%%%%%%%%%%%%%%%%%%%%%%%%%%%%%%%%%%%%%%%%%%%%%%%%%%%%%%%%%%%%%%%%%%%%%
\subsection{Background material on massive and strictly/partially massless integer-spin theories on $dS_{4}$}
Before delving into the study of the (unconventional) conformal invariance of maximal depth partially massless fields, let us review the basics concerning massive and strictly/partially massless (classical) fields corresponding to Eqs.~(\ref{EOM massive}).

As was discovered by Higuchi~\cite{Higuchiforb, STSHS, Yale_Thesis}, the value of the mass parameter in Eqs.~(\ref{EOM massive}) determines whether the field theory is massive or strictly/partially massless, as well as whether the theory is unitary - i.e. whether the $so(4,1)$ representation realised on the solution space is unitary.\footnote{By unitarity of the representation we mean that there exists a positive-definite, dS invariant scalar product for the mode solutions of the field equations~(\ref{EOM massive}). By `massive' we mean that the theory does not enjoy a gauge symmetry.} In particular, we have non-unitarity for $$m^{2} < (s-1)s,$$ while unitary massive theories\footnote{A massive theory has $2s+1$ propagating degrees of freedom~\cite{Deser_Waldron_phases}.} obey the `Higuchi bound'
\begin{align}
    m^{2} > (s-1)s.
\end{align}
For each of the following special tunings of $m^{2}$~\cite{Higuchiforb, STSHS, Yale_Thesis}:
\begin{align}\label{pm_tuning_bosons}
    m^{2}=(\uptau -1)(2s-\uptau),\hspace{10mm}(\uptau=1,...,s),
\end{align}
the $so(4,1)$ representation is unitary, while the field theory enjoys a gauge symmetry, i.e. $h_{\mu_{1} ... \mu_{s}}$ is a gauge potential. 
A gauge field $h_{\mu_{1} ... \mu_{s}}$ with mass parameter given by~(\ref{pm_tuning_bosons}) is called partially massless field of depth $\uptau$~\cite{Deser_Waldron_phases, Deser_Waldron_null_propagation, Deser_Waldron_partial_masslessness}. The value $\uptau =1$, for which $m^{2}=0$, corresponds to the strictly massless spin-$s$ field with two propagating helicity degrees of freedom, $\pm s$. The values $ \uptau = 2, ..., s$ correspond to partially massless spin-$s$ fields with $2 \uptau$ propagating helicity degrees of freedom: $(\pm s, \pm (s-1), ..., \pm (s-\uptau+1))$~\cite{Deser_Waldron_phases, Deser_Waldron_null_propagation, Deser_Waldron_partial_masslessness}. 

The infinitesimal on-shell gauge transformations for the strictly/partially massless spin-$s$ fields of depth $\uptau$ are of the form~\cite{STSHS, Deser_Waldron_partial_masslessness}
\begin{align}\label{gauge transformations}
   \delta h_{\mu_{1} ... \mu_{s}} =\nabla_{(\mu_{1}} \nabla_{\mu_{2}} ... \nabla_{\mu_{\uptau}} \alpha_{\mu_{\uptau +1} ... \mu_{s})}+..., 
\end{align}
where $\alpha_{\mu_{\uptau+1}...\mu_{s} }$ is a totally symmetric tensor gauge function of rank-$(s - \uptau)$. In `$...$', there are terms corresponding to completely symmetrised products of the metric tensor with covariant derivatives acting on $\alpha_{\mu_{\uptau+1}...\mu_{s} }$~\cite{STSHS, Deser_Waldron_phases, Deser_Waldron_null_propagation, Deser_Waldron_partial_masslessness}. The gauge transformations~(\ref{gauge transformations}) leave Eqs.~(\ref{EOM massive}) invariant - given that the gauge functions satisfy certain extra conditions~\cite{STSHS}. For example, the on-shell gauge transformations for strictly massless (i.e., $\uptau = 1$) spin-$s$ fields are~\cite{STSHS}
\begin{align}\label{onshell gauge transformations strictly massless any spin}
     \delta h_{\mu_{1} ... \mu_{s}} = \nabla_{(\mu_{1}}a_{\mu_{2}...\mu_{s})} , 
\end{align}
with
\begin{align}\label{onshell gauge functions strictly massless any spin}
     &\Box a_{\mu_{2}...\mu_{s}} = -(s^{2}-1) ~a_{\mu_{2}...\mu_{s}}, \nonumber \\
     &\nabla^{\mu_{2}} a_{\mu_{2} \mu_{3}...\mu_{s}} = 0 ,~ ~g^{\mu_{2}   \mu_{3}} a_{\mu_{2}\mu_{3}...\mu_{s}} = 0.
 \end{align}

Note that maximal depth (i.e. $\uptau = s$) partially massless fields have a scalar gauge function in their corresponding gauge transformation~(\ref{gauge transformations})\footnote{Of course, the Maxwell theory, i.e. the strictly massless gauge potential $h_{\mu_{1}}$ ($\uptau = s = 1$) also enjoys gauge symmetry with scalar gauge functions ($\delta h_{\mu_{1}} = \nabla_{\mu_{1}}a$).}. For example, in the case of the partially massless graviton ($\uptau = s =2$) we have the following on-shell gauge transformation~\cite{DESER_NEPOM_1, DESER_NEPOM_2}:
\begin{align}\label{gauge transformations p.m. graviton}
     \delta h_{\mu_{1} \mu_{2}} = \left( \nabla_{(\mu_{1}}  \nabla_{\mu_{2})} +g_{\mu_{1}   \mu_{2}}   \right)a , 
\end{align}
with
\begin{align}\label{gauge function onshell p.m. graviton}
     &\Box \,a = -4\,a.
 \end{align}
The action functional of the partially massless graviton is known to be invariant under off-shell gauge transformations of the form~(\ref{gauge transformations p.m. graviton}) with $\alpha$ being a generic scalar gauge function. [In the case of off-shell gauge transformations the gauge function $\alpha$ does not have to satisfy Eq.~(\ref{gauge function onshell p.m. graviton}) - see, e.g., Ref.~\cite{de_Rham}, as well as Section~\ref{sec_pm graviton}].

%%%%%%%%%%%%%%%%%%%%%%%%%%%%%%%%%%%%%%%%%%%%%%%%%%%%%%%%%%%%%%%%%%%%%%%%%%%%%%%%%%%%%%%%%%5
The Unitary Irreducible Representations (UIRs) of the dS algebra $so(D,1)$ corresponding to totally symmetric massive and strictly/partially massless integer-spin fields have been discussed in detail in Ref.~\cite{Yale_Thesis}. For more recent discussions on dS representation theory and ``field theory - representation theory dictionaries'' see Refs.~~\cite{Mixed_Symmetry_dS, Sun, Gizem, Letsios_announce, Alan, Letsios_in_progress, Schaub, SchaubWalk}.

%%%%%%%%%%%%%%%%%%%%%%%%%%%%%%%%%%%%%%%%%%%%%%%%%%%%%%%%%%%%%%%%%%%%%%%%%%%%%%%%%%%%%%%%%%%%%%%%%%%%%%%%%%%%%%%%%%%%%%%%%%%%%%%%%%%%%%%%%%%%%%%%%%%%%%%%%%%%%%%%%%%%%%%%%%%%
\subsection{What is (not) known about the conformal symmetry of partially massless fields of maximal depth? Does invariance under transformations generated by conformal Killing vectors not imply $so(4,2)$ symmetry?}

%%%%%%%%%%%%%%%%%%%%%%%%%%%%%%%%%%%%%%%%%%%%%%%%%%%%%%%%%%%%%%%%%%%%%%%%%%%%%%%%

 Partially massless integer-spin fields of maximal depth correspond to the value $\uptau = s$ in Eq.~(\ref{pm_tuning_bosons}), which means that their mass parameters are given by:
\begin{align} \label{max depth value}
    m^{2} = s(s-1).
\end{align}
{In the transverse-traceless gauge the field equations~(\ref{EOM massive}) are}~\footnote{The field equations~(\ref{EOM max depth}) with $s=1$ correspond to the strictly massless vector field (Maxwell theory), while for $s=0$ they essentially reduce to the field equation for the conformally coupled scalar $\nabla^{\alpha} \nabla_{\alpha} \Phi = 2 \Phi$.}
\begin{align} \label{EOM max depth}
    &\Big( \nabla^{\alpha} \nabla_{\alpha}-(s+2) \Big)h_{\mu_{1}...\mu_{s}} = 0, \nonumber\\
    & \nabla^{\alpha} h_{\alpha \mu_{2}...\mu_{s}}=0, \hspace{4mm} g^{\alpha \beta} h_{\alpha \beta \mu_{3}...\mu_{s}}=0,
\end{align}
{where the gauge conditions satisfied by the fields (gauge potentials) $h_{\mu_{1}...\mu_{s}}$ are given in the second line. In particular, these are the conditions of transversality ($\nabla^{\alpha} h_{\alpha \mu_{2}...\mu_{s}}=0$) and tracelessness ($g^{\alpha \beta} h_{\alpha \beta \mu_{3}...\mu_{s}}=0$). Note that in the massive case~(\ref{EOM massive}) [i.e. when the mass parameter is not tuned to a strictly/partially massless value], the conditions of transversality and tracelessness are consequences of the equations of motion, while in the strictly/partially massless cases these conditions are a gauge choice.}

\noindent \textbf{What is known about the conformal invariance of maximal depth partially massless fields.} A maximal depth partially massless field, $h_{\mu_{1}...\mu_{s}}$ ($s \geq 2$), shares similarities with the Maxwell field~\cite{Deser_Waldron_Conformal} (i.e., with the strictly massless spin-1 field which corresponds to the value $s=1$ in Eq.~(\ref{EOM max depth})). In particular, they both propagate on the lightcone\footnote{Lightcone propagation is a feature of strictly and partially massless fields of any depth on $dS_{4}$~\cite{Deser_Waldron_null_propagation}.}, while also their gauge transformations~(\ref{gauge transformations}) include scalar gauge functions. Moreover, the results of Ref.~\cite{Deser_Waldron_Conformal} indicate that, as in the Maxwell case, the maximal depth partially massless theories enjoy invariance not only under infinitesimal dS transformations but also under infinitesimal transformations generated by the five genuine conformal Killing vectors of $dS_{4}$.
As we will explain shortly, in the case of maximal depth partially massless fields, there are still open questions concerning the explicit form of the conformal transformations, as well as the structure of the symmetry algebra.

\noindent  \textbf{Note on terminology.} By `genuine conformal Killing vectors' we mean the five non-Killing conformal Killing vectors of $dS_{4}$ (see Eq.~(\ref{CKV equation})). %By `conformal transformations' we mean the transformations generated by the genuine conformal Killing vectors.
\\
\\
\noindent \textbf{Reviewing the conformal invariance of the Maxwell field.} To demonstrate what is missing in our current understanding of the conformal symmetry of maximal depth partially massless theories, let us briefly review the well-known infinitesimal conformal invariance of the Maxwell theory on $dS_{4}$. The Maxwell action on $dS_{4}$\footnote{The determinant of the dS metric is denoted as $g$.}, 
\begin{align}\label{Maxwell action}
 S_{1} = -\frac{1}{4}  \int \sqrt{-g} ~d^{4}x~ F^{\mu \nu}F_{\mu \nu},\hspace{5mm} F_{\mu \nu}  \equiv \nabla_{\mu}h_{\nu}   - \nabla_{\nu}h_{\mu} ,
\end{align}
is known to be invariant under the ten dS isometries, as well as under the five conformal isometries generated by genuine conformal Killing vectors of $dS_{4}$. {Let us briefly discuss the well-known fact that the invariance of the Maxwell theory under conformal isometries on a fixed background spacetime arises from the combination of the more general diffeomorphism and  Weyl invariances of the theory in curved spacetime.\footnote{See e.g. Section IIA in \cite{Stergiou} for a pedagogical review on the relation between Weyl and conformal invariance.} Consider the Maxwell action on an arbitrary 4-dimensional curved spacetime with metric $g_{\mu \nu}'$ 
\begin{align*}
 S_{1}[h, g'] = -\frac{1}{4}  \int \sqrt{-g'} ~d^{4}x~ g'_{\mu \alpha}  \,g'_{\nu \beta} F^{\alpha \beta} F^{\mu \nu}.
\end{align*}
 The action is invariant under infinitesimal diffeomorphisms 
 \begin{align*}
     \delta^{diff}_{K}g'_{\mu \nu} = 2 \nabla_{(\mu} K_{\nu)},~~~ ~~\delta^{diff}_{K}h_{\mu} = \mathcal{L}_{K} h_{\mu},
 \end{align*} 
 where $K$ is a generic vector field. Moreover, the action is invariant under Weyl rescalings, which we present in their infinitesimal form,
 \begin{align*}
     \delta^{w} g'_{\mu \nu} = 2 \, \sigma  \,  g'_{\mu \nu}, ~ ~ ~ ~\delta^{w}h_{\mu} = 0.
 \end{align*}
 These imply $\delta \sqrt{-g'} = 4 \sigma \sqrt{-g'} $, $\delta^{w}F^{\alpha \beta} = - 4 \, \sigma \, F^{\alpha \beta}$ and $\delta^{w}F_{\alpha \beta} = 0$. Now let us fix a background metric and, for simplicity, let this be the background dS metric $g'_{\mu \nu} = g_{\mu \nu}$. The Maxwell theory is now invariant under transformations that combine diffeomorphisms and Weyl rescalings such that the background metric $g_{\mu \nu}$ is preserved. These are just the infinitesimal conformal isometries of  $g_{\mu \nu}$, $$ \delta^{conf}_{K}g_{\mu \nu} \equiv  \delta^{w} g_{\mu \nu} + \delta^{diff}_{K}g_{\mu \nu} = 0,$$ and it follows that $\sigma = - \tfrac{1}{4}\nabla^{\alpha} K_{\alpha}$, where $K^{\mu}$ is any of the fifteen conformal Killing vectors of $dS_{4}$ (10 Killing plus 5 genuine conformal Killing vectors). The Maxwell gauge potential transforms under infinitesimal conformal isometries as 
 $$ \delta^{conf}_{K}h_{\mu \nu}  =  \delta^{w} h_{\mu} + \delta^{diff}_{K}h_{ \mu} = \mathcal{L}_{K}h_{\mu}.$$
 These transformations leave the Maxwell action~(\ref{Maxwell action}) invariant.}

{Let us now focus on the 5 genuine infinitesimal conformal isometries of the Maxwell theory as these are the ones that will generalise in an unconventional way in the case of maximal depth partially massless fields.} From the discussion in the previous paragraph it follows that the Maxwell action on $dS_{4}$ remains invariant under the following transformations:
\begin{align}
    \delta_{V} h_{\mu} &= V^{\rho}F_{\rho \mu}  \label{conf transf Maxwell improved 1}\\
    &= \mathcal{L}_{V}h_{\mu} + \nabla_{\mu} \left(-V_{\rho}h^{\rho}    \right),\label{conf transf Maxwell improved 2}
\end{align}
where $V^{\rho}$ is any of the five genuine conformal Killing vectors of $dS_{4}$ (or a linear combination thereof). {Here we have chosen to work with the ``improved'' infinitesimal conformal transformations~(\ref{conf transf Maxwell improved 1}) for later convenience - these are off-shell symmetries of the action~(\ref{Maxwell action}), as well as on-shell symmetries of the field equations in the Lorenz gauge~(\ref{EOM max depth}).\footnote{The conventional conformal transformations of the Maxwell gauge potential $\mathcal{L}_{V}h_{\mu}$ are not symmetries of the field equations~(\ref{EOM max depth}) in the Lorenz gauge. In Eq.~(\ref{conf transf Maxwell improved 2}), we have expressed $\delta_{V}  h_{\mu}$ as a sum of two off-shell symmetry transformations of the action: a conventional infinitesimal conformal transformation, $\mathcal{L}_{V}  h_{\mu}$, plus a gauge transformation, $\nabla_{\mu} (-V_{\rho}h^{\rho})$.} As for the symmetry algebra of the theory, by calculating the commutators of the corresponding differential operators describing the infinitesimal field transformations (i.e., $[\delta_{V}, \delta_{V'} ]h_{\mu}$, $[\mathcal{L}_{\xi} ,\mathcal{L}_{\xi'}]h_{\mu}$, $[\mathcal{L}_{\xi} , \delta_{V}] h_{\mu}$), it is easy to verify that the five ``improved'' conformal transformations $\delta_{V} h_{\mu}$, together with the ten infinitesimal dS transformations~$\mathcal{L}_{\xi}h_{\mu}$~[Eq.~(\ref{Lie derivative})], generate a representation of the conformal algebra $so(4,2)$~\footnote{To be precise, when using the ``improved'' conformal transformations $\delta_{V}h_{\mu}$, instead of the conventional ones $\mathcal{L}_{V}h_{\mu}$, the algebra generated by the five transformations $\delta_{V}h_{\mu}$ and the ten dS transformations~(\ref{Lie derivative}) closes on $so(4,2)$ up to gauge transformations.}.
\\
\\
\noindent \textbf{What is \textit{not} known about the conformal invariance of maximal depth partially massless fields.} Let us now shift focus again to the maximal depth partially massless fields and see what is missing in our understanding (by comparing with the Maxwell case presented above). In Ref.~\cite{Deser_Waldron_Conformal}, Deser and Waldron showed that the first-order actions (expressed in terms of the propagating degrees of freedom) for maximal depth partially massless spin-$s \geq 2$ fields are invariant under infinitesimal dS dilations. Then, the commutators between infinitesimal dS transformations and dS dilations also imply that the aforementioned first-order actions are invariant under infinitesimal transformations generated by any of the five genuine conformal Killing vectors of $dS_{4}$. The authors of Ref.~\cite{Deser_Waldron_Conformal} worked non-covariantly (in the steady state coordinates of $dS_{4}$) and they ingeniously exploited the similarity between the aforementioned first-order actions and the first-order action for the conformally coupled scalar field.
%%%%%%%%%%%%%%%%%%%%%%%%%%%%%%%%%%%%%%%%%%%%%%%%%%%%%%%%%%%%%%%%%%%%%%%%%%%%%%%%%%%%%%%%%%%%%%%%%%%%%%%%%%%%%%%%%%%%%%%%%%%%%%%%%%%%%%%%
\textit{\textbf{However, covariant expressions [such as Eqs.~(\ref{conf transf Maxwell improved 1}) and (\ref{conf transf Maxwell improved 2}) in the Maxwell case] for generic infinitesimal conformal transformations acting on the fields $h_{\mu_{1} \mu_{2}...\mu_{s}}$, leaving Eqs.~(\ref{EOM max depth}) (or the covariant actions) invariant, were not presented in Ref.~\cite{Deser_Waldron_Conformal}. In fact, such covariant expressions are absent from the literature. We present them here for the first time at the level of the field equations~(\ref{EOM max depth}) for spin-$s \geq 2$, and at the level of the action for $s=2$.}} Interestingly, we find that the infinitesimal conformal transformations for maximal depth partially massless higher-spin fields do \textbf{not} just correspond to a Lie derivative with respect to genuine conformal Killing vectors plus a conformal weight term, as one might expect. The expressions~[(\ref{unvonv conf pm graviton}) and (\ref{unvonv conf pm spin-s>2})] turn out to be similar in form to the ``improved'' conformal transformations~(\ref{conf transf Maxwell improved 1}) of the Maxwell theory, but, unlike the Maxwell case, they cannot be expressed as a sum of a conventional conformal transformation plus a gauge transformation (as the latter contains second order derivatives).
\\
\\
\noindent \textbf{Does invariance under transformations generated by genuine conformal Killing vectors not imply $so(4,2)$ symmetry?} According to the representation-theoretic study of Ref.~\cite{Conf_Bekaert}, the maximal depth partially massless fields on $dS_{4}$ do \textbf{not} enjoy $so(4,2)$ symmetry. In the present article, our findings seem to verify this claim. To be specific, we will show that, unlike the Maxwell case, the commutator between two infinitesimal conformal transformations of maximal depth partially massless fields is not equal to an infinitesimal dS transformation. {This means that the algebra does not close on $so(4,2)$, while the question of whether the algebra closes in general is left open in this paper - see Sections~\ref{sec_pm graviton} and \ref{sec_mac depth pm spin s>2}. In particular, the commutator between two infinitesimal conformal transformations, generated by two genuine conformal Killing vectors $V$ and $V'$, respectively, is equal to a new symmetry transformation that is second-order in derivatives (the algebra would close on $so(4,2)$ if this commutator were equal to an infinitesimal dS isometry transformation, namely a Lie derivative with respect to the dS Killing vector $[V,V']$). The new second-order symmetry transformation~(\ref{higher symmetry pm graviton}) does not seem to correspond to a gauge transformation, and it does \textbf{not} reduce to an infinitesimal dS isometry (Lie derivative) on-shell - see Subsection \ref{subsec_algebra pm graviton}}. We will thus call our conformal transformations `unconventional conformal transformations'. 

The main results and investigations of the present article are summarised in the next Subsection. `Conformal-like' symmetries for strictly massless higher-spin bosons and fermions on $dS_{4}$ have been uncovered in Refs.~\cite{Letsios_Hidden} and \cite{Letsios_Higuchi_Hidden}, respectively.

%%%%%%%%%%%%%%%%%%%%%%%%%%%%%%%%%%%%5

%%%%%%%%%%%%%%%%%%%%%%%%%%%%%%%%%%%%%%%%%%%%%%%%%%%%%%%%%%%%%%%%%%%%%%%%%%%%%%%%%%%%%%%%%%%%%5
\subsection{Outline and main results: unconventional conformal symmetry and its role in a complex partially massless supermultiplet}
The main results and outline of the present article are presented in the following list:
\begin{itemize}
    \item Section~\ref{sec_pm graviton} concerns the free partially massless graviton on $dS_{4}$. In Subsection~\ref{subsec_pm grav action invariance}, we uncover the explicit covariant form of the `unconventional' (infinitesimal) conformal transformation~(\ref{unvonv conf pm graviton}) and we show that it leaves the action invariant. (The reason why we call this symmetry `unconventional' has been explained in the previous Subsection.) The structure of the symmetry algebra of the theory is studied in Subsection~\ref{subsec_algebra pm graviton}. We find that the algebra does not close on $so(4,2)$ (verifying the claims of Ref.~\cite{Conf_Bekaert}), due to the appearance of a new symmetry transformation~(\ref{higher symmetry pm graviton}) that is second-order in derivatives (this symmetry appears in the commutator between two unconventional conformal transformations). Our results concerning the closure of the algebra are inconclusive - however, it is clear that there is no closure on $so(4,2)$.

    \item Section~\ref{sec_mac depth pm spin s>2} concerns the free spin-$s \geq 2$ partially massless fields of maximal depth on $dS_{4}$. In Subsection~\ref{subsec_unconf conf spin s>=2}, we uncover the covariant expressions for the `unconventional' (infinitesimal) conformal transformations~(\ref{unvonv conf pm spin-s>2}) that leave the field equations~(\ref{EOM max depth}) invariant. The structure of the symmetry algebra is studied in Subsection~\ref{subsec_algebra pm max depth spin s>2} and, as in the spin-2 case, we find that it does not close on $so(4,2)$.

    \item In Section~\ref{sec_susy}, we uncover a new non-interacting complex supermultiplet containing a complex partially massless spin-2 field (complex partially massless graviton)~(\ref{EOM pm complex graviton}) and a complex spin-3/2 field with zero mass parameter~(\ref{EOM spin-3/2}). Our analysis of this supersymmetric theory is not extensive, and our main objective is simple and twofold. First, we demonstrate, for the first time, the existence of a supermultiplet containing partially massless fields on $dS_{4}$. In this theory, the commutator of two supersymmetry (SUSY) variations does not close either on the dS algebra $so(4,1)$ or on the conformal algebra $so(4,2)$. Second, we showcase the appearance of the unconventional conformal symmetries in the even subalgebra of our supermultiplet. Specifically, we present the SUSY transformations that transform the classical solution spaces of the bosonic and fermionic field equations into each other~(i.e. we demonstrate the realisation of a SUSY representation). Then, we calculate the commutator~(\ref{[susy, susy]pm grav compl}) of two SUSY transformations acting on the partially massless spin-2 field and we find two even symmetries: an unconventional conformal symmetry and a symmetry with second-order derivatives. Although our results demonstrate that the even algebra does \textbf{not} close on $so(4,2)$, its full structure is still unknown.

{Our novel supermultiplet could be related to a complex version of partially massless dS supergravity (if such a non-linear theory exists). We note that the idea to look for dS supergravity, as well as the suggestion to drop the Majorana condition, was given by Deser and Waldron in \cite{Deser_Waldron_ArbitrarySR}, but it was not developed in detail.}

    \item In Section~\ref{sec_discussion}, we present the open questions that arise from our findings and we discuss future research directions.
\end{itemize}

 %\item We briefly discuss the massless spin-1/2 field on $dS_{4}$, which is known to be invariant under the familiar 15-dimensional conformal algebra $so(4,2)$. We show that the theory also enjoys fifteen conformal-like symmetries. Each conformal-like symmetry corresponds to the product of an axial rotation times a familiar infinitesimal conformal ($so(4,2)$) transformation. Then, we show that the symmetry algebra generated by the familiar conformal transformations together with the conformal-like transformations is isomorphic to $so(4,2) \bigoplus so(4,2)$. However, this is a chiral symmetry, i.e. $so(4,2)^{+} \bigoplus so(4,2)^{-}$, where $so(4,2)^{\pm}$ acts non-trivially on mode solutions with chirality $\pm 1/2$, and trivially on mode solutions with chirality $\mp 1/2$. In other words, a massless spin-1/2 field with a single, fixed, chirality has only fifteen non-trivial symmetries, corresponding to familiar $so(4,2)$ conformal transformations. But a massless theory with both chiralities enjoys a $so(4,2)^{+} \bigoplus so(4,2)^{-}$ chiral symmetry.

%%%%%%%%%%%%%%%%%%%%%%%%%%%%%%%%%%
\subsection{Notation and conventions}

 We use the mostly plus metric sign convention for $dS_{4}$. Lowercase Greek tensor indices refer to components with respect to the `coordinate basis' on $dS_{4}$. Lowercase Latin tensor indices refer to components with respect to the vielbein basis. Repeated indices are {summed} over. We denote the symmetrisation of indices with the use of round brackets, e.g. $A_{(\mu \nu)} \equiv  (A_{\mu \nu}+A_{\nu \mu})/2$, and the anti-symmetrisation with the use of square brackets, e.g. $A_{[\mu \nu]} \equiv  (A_{\mu \nu}-A_{\nu \mu})/2$. We suppress spinor indices. By `genuine conformal Killing vectors' we mean the five non-Killing conformal Killing vectors of $dS_{4}$ with non-vanishing divergence - see Eqs.~(\ref{CKV equation}), (\ref{V=nabla phi}). The commutator of covariant derivatives for a totally symmetric tensor $h_{\mu_{1}...\mu_{s}}$ is
\begin{align}\label{commut of cov derivs}
 [\nabla_{\mu} , \nabla_{\nu}]h_{\mu_{1}...\mu_{s}}  =~s\, \Big( {g}_{\mu(\mu_{1}} h_{\mu_{2}...\mu_{s})\nu}   -{g}_{\nu(\mu_{1}} h_{\mu_{2}...\mu_{s})\mu} \Big) .
\end{align}
Our conventions for fermionic fields and gamma matrices on $dS_{4}$ are those of Refs.~\cite{Letsios_Hidden, Letsios_announce}.

%%%%%%%%%%%%%%%%%%%%%%%%%%%%%%%

%%%%%%%%%%%%%%%%%%%%%%%%%%%%%%%%%%%%%%%%%%%%%%%%%%%%%%%%%%%%%%%%%%%%%%%

%%%%%%%%%%%%%%%%%%%%%%%%%%%%%%%%%%%%%%%%%%%%%%%%%%%%%%%%%%%%%%%%%%%%%%%%%%%%%
\section{Unconventional conformal symmetry of the covariant action for the partially massless graviton}\label{sec_pm graviton}

\subsection{Conformal Killing vectors of $dS_{4}$} \label{subsec_conf killing vectors}
Four-dimensional de Sitter spacetime has ten Killing vectors~(\ref{KV equation}) and five genuine conformal Killing vectors. The latter satisfy 
\begin{align}\label{CKV equation}
  \nabla_{\mu}  V_{\nu} + \nabla_{\nu}  V_{\mu}= g_{\mu \nu}\frac{\nabla^{\alpha}    V_{\alpha}}{2},
\end{align}
where $\nabla^{\alpha}    V_{\alpha} \neq 0$. 
Each of the five conformal Killing vectors of $dS_{4}$ can be expressed as:
\begin{align}\label{V=nabla phi}
    V_{\mu}= \nabla_{\mu} \phi_{V},
\end{align}
where the scalar function $\phi_{V}$ satisfies 
{\begin{align}
   &\nabla_{\mu}V_{\nu}=~ \nabla_{\mu} \nabla_{\nu}\phi_{V} =  - g_{\mu \nu}\phi_{V}\label{properties of phi 1} 
\end{align}}
(see, e.g. Refs.~\cite{Allen} and \cite{Letsios_Hidden}).

The Killing and genuine conformal Killing vectors of $dS_{4}$ generate the algebra $so(4,2)$ with commutation relations:
\begin{align}
   & [\xi, \xi'  ]^{\mu} = \mathcal{L}_{\xi} \xi^{'\mu}, \\
    & [\xi, V  ]^{\mu} = \mathcal{L}_{\xi} V^{\mu}, \\
     & [V, V'  ]^{\mu} = \mathcal{L}_{V} V^{'\mu},
\end{align}
where $\xi , \xi'$ are any two Killing vectors, $V$ and $V'$ are any two genuine conformal Killing vectors, $\mathcal{L}_{\xi}V^{\mu}$ is a genuine conformal Killing vector, while $\mathcal{L}_{V}V^{'\mu}$ is a Killing vector.

%%%%%%%%%%%%%%%%%%%%%%%%%%%%%%%%%%%%%%%%%%%%%%%%%%%%%%%%%%%%%%%%%%%%%%%%%%%%%%%%%%%%%%%%%%%%%%%%%%%%%%%%%%%%%%%%%%%%%%%
\subsection{Unconventional conformal symmetry transformations and invariance of the action} \label{subsec_pm grav action invariance}
The standard covariant action~\cite{Higuchiforb} for the partially massless graviton can be equivalently expressed {as}~\cite{spin2EM}
\begin{align}\label{pm graviton action}
  S_{2}= - \frac{1}{4}   \int d^{4}x \, \sqrt{-g} \, \left(  F_{\mu \nu | \rho} F^{\mu \nu | \rho} - 2 \, g^{\alpha \beta}\, F_{\mu \alpha | \beta} ~g^{\gamma \delta} F^{\mu}_{\hspace{2mm} \gamma  |  \delta}  \right) , 
\end{align}
where 
\begin{align} \label{pm spin-2 field strength}
    F_{\mu \nu | \rho} = F_{[\mu \nu] | \rho} \equiv \nabla_{\mu}  h_{\nu \rho}  - \nabla_{\nu} h_{\mu \rho},
\end{align}
is the field strength (the symbol `$|$' has been used to distinguish between the anti-symmetric pair of indices $\mu,\nu$ and the third index $\rho$). The field strength, as well as the action~(\ref{pm graviton action}), are invariant under off-shell gauge transformations $$\delta h_{\nu \rho}= \left( \nabla_{(\nu}    \nabla_{\rho)}+g_{\nu \rho} \right) a.$$
The field strength also satisfies the following off-shell identities:
\begin{align}
   & F_{\mu \nu | \rho}+ F_{ \rho\mu |\nu }+F_{ \nu \rho |\mu }=0, \\
  & \nabla_{[\kappa}  F_{\mu \nu] | \rho}=0.
\end{align}
For later convenience, integrating by parts and using Eq.~(\ref{commut of cov derivs}) (and dropping the total divergence terms), we can re-write the action as
\begin{align}\label{pm graviton action simple form}
    S_{2}=\frac{1}{2}  \int d^{4}x \, \sqrt{-g} h^{\mu \nu}H_{\mu \nu},
\end{align}
 where we have defined
\begin{align}
 H_{\mu \nu} =&-g^{\beta \gamma} \nabla^{\alpha}     F_{\alpha  \beta| \gamma } \, g_{\mu \nu}+ \nabla^{\alpha}F_{\alpha       \mu |  \nu} + \nabla_{\nu}F_{\mu \beta| \gamma}\, g^{\beta \gamma}  \\
 =& - 2\nabla_{(\mu} \nabla^{\alpha}h_{\nu)  \alpha}+ \Box h_{\mu \nu}-g_{\mu \nu}\,\Box h^{\alpha}_{\alpha} +\nabla_{\mu} \nabla_{\nu}h^{\alpha}_{\alpha} \nonumber \\
    &+ g_{\mu \nu}\, \nabla^{\alpha}\nabla^{\beta}h_{\alpha \beta} -4\, h_{\mu \nu}+g_{\mu \nu}h^{\alpha}_{\alpha}.
\end{align}
It is easy to show that the last two terms in the first line satisfy
$$ \nabla^{\alpha}F_{\alpha       \mu |  \nu} + \nabla_{\nu}F_{\mu \beta| \gamma}\, g^{\beta \gamma} = \nabla^{\alpha}F_{\alpha       (\mu |  \nu)} + \nabla_{(\nu}F_{\mu) \beta| \gamma}\, g^{\beta \gamma} .  $$

\noindent    \textbf{Unconventional conformal transformations.} The unconventional infinitesimal conformal transformations of the partially massless graviton are given by
\begin{align} \label{unvonv conf pm graviton}
    \delta_{V} h_{\mu \nu } =V^{\rho} F_{\rho (\mu | \nu)} = V^{\rho}  \left( \nabla_{\rho}h_{\mu \nu}- \nabla_{(\mu}h_{\nu)  \rho}     \right),
\end{align}
where $V^{\rho}$ is any genuine conformal Killing vector~(\ref{CKV equation}) of $dS_{4}$. Note that, unlike the Maxwell case~(\ref{conf transf Maxwell improved 1}), the transformation~(\ref{unvonv conf pm graviton}) has an unconventional form, as it cannot be expressed as:  $\mathcal{L}_{V} h_{\mu \nu}$ $+(\text{conformal weight term})$+(gauge transformation). Also, as we will see in the following Subsection, if one considers on-shell partially massless gravitons, it happens that the transformations~(\ref{unvonv conf pm graviton}) preserve the transverse-traceless conditions.

%%%%%%%%%%%%%%%%%%%%%%%%%%%%%%%%%%%%%%%%%%%%%%%%%%%%%%%%%%%%%%%%%%%%%%%%%%%%%%%%%%%%%%%%%%%%%%%%%%%%%%%%%%%%%%%%%%%%%%%

\noindent   \textbf{Invariance of the covariant action.}
After a straightforward off-shell calculation, we find that the field strength transforms under~(\ref{unvonv conf pm graviton}) as
\begin{align}\label{unvonv conf pm graviton fieldstr}
\delta_{V} F_{\alpha \mu |\nu} & \equiv  \nabla_{\alpha} \, \delta_{V} h_{\mu \nu}  - \nabla_{\mu}\, \delta_{V} h_{\alpha \nu} \nonumber \\
&= V^{\rho}\left(\nabla_{\rho}F_{\alpha \mu | \nu}-\frac{1}{2}\nabla_{\nu}F_{\alpha \mu | \rho}\right)-\frac{3}{2} \phi_{V}F_{\alpha \mu | \nu}.
\end{align}
Using this expression, and performing a lengthy calculation, we find
\begin{align}\label{transformation of Hmunu}
\delta_{V} H_{\mu \nu}  \equiv &~-g^{\beta \gamma} \nabla^{\alpha}    \, \delta_{V}F_{\alpha  \beta| \gamma } \, g_{\mu \nu}+ \nabla^{\alpha}\,\delta_{V}F_{\alpha       (\mu |  \nu)} + \nabla_{(\nu}\,\delta_{V}F_{\mu) \beta| \gamma}\, g^{\beta \gamma}  \nonumber \\
 =& V^{\rho}  \nabla_{\rho}H_{\mu \nu}   -V_{(\mu}   
 \nabla^{\lambda}H_{\nu)  \lambda}  - 3 \phi_{V}H_{\mu \nu}.
\end{align}
Then, it is straightforward to show that the variation of the action~(\ref{pm graviton action simple form}),
\begin{align}
\delta_{V}    S_{2}=\frac{1}{2}  \int d^{4}x \, \sqrt{-g} \left(  \delta_{V} h^{\mu \nu}\,H_{\mu \nu}  +   h^{\mu \nu}\, \delta_{V}H_{\mu \nu} \right),
\end{align}
is a total divergence. In other words, the unconventional conformal transformation~(\ref{unvonv conf pm graviton}) is an off-shell symmetry of the action.

\noindent  \textbf{Invariance of the field equations~(\ref{EOM max depth}) for $\bm{s=2}$.} It is easy to show that the unconventional conformal transformation~(\ref{unvonv conf pm graviton}) is also a symmetry of the field equations~(\ref{EOM max depth}) in the transverse-traceless gauge (for $s=2$). In other words, if $h_{\mu \nu}$ is a solution of Eqs.~(\ref{EOM max depth}), then so is $\delta_{V} h_{\mu \nu}$. See the next Section for some details concerning the calculations for all spins $s \geq 2$.
%%%%%%%%%%%%%%%%%%%%%%%%%%%%%%%%%%%%%%%%%%%%%%%%%%%%%%%%%%%%%%%%%%%%%%%%%%%%%%%%%%%%%%%%%%%%%%%%%%%%%%%%%%%%%%%%%%%%%%%%
\subsection{The algebra does not close on \texorpdfstring{$so(4,2)$}{so(4,2)}} \label{subsec_algebra pm graviton}
We now know that the partially massless graviton has fifteen infinitesimal symmetry generators: ten dS transformations~(\ref{Lie derivative}) and five unconventional conformal transformations~(\ref{unvonv conf pm graviton}). Let us calculate the corresponding commutators in order to (try to) understand the structure of the symmetry algebra.
We find the following results:
\begin{align}
    [\mathcal{L}_{\xi},  \mathcal{L}_{\xi '} ] h_{\mu \nu} &= \mathcal{L}_{ [\xi, \xi']}h_{\mu \nu}, \label{[KV,KV]pm graviton} \\
    [\mathcal{L}_{\xi},  \delta_{V} ] h_{\mu \nu}=  & ~ \mathcal{L}_{\xi} \, \delta_{V}h_{\mu \nu}- \delta_{V}  \mathcal{L}_{\xi} h_{\mu \nu}\nonumber   \label{[KV,conf] pm graviton}  \\
    =& ~\delta_{[\xi ,V ]}h_{\mu \nu}, \\
    [\delta_{V},  \delta_{V'} ] h_{\mu \nu}=  & ~ \delta_{V} \, \delta_{V'}h_{\mu \nu}- \delta_{V'} \, \delta_{V}h_{\mu \nu}\nonumber\\
    =& ~ \frac{3}{2}  \left(  [V,V']^{\rho}F_{\rho (\mu|\nu)} +\frac{1}{6}  \nabla_{\alpha}[V,V']_{\beta}\,\nabla_{(\mu} F^{\alpha \beta}_{\hspace{5mm}| \nu)} \right) \label{[conf,conf]pm graviton},
\end{align}
where $\xi ,\xi'$ are any two dS Killing vectors, $V$ and $V'$ are any two dS genuine conformal Killing vectors, while $[V, V']^{\rho} = \mathcal{L}_{V}V^{' \, \rho}$ is a dS Killing vector. The commutation relations~(\ref{[KV,KV]pm graviton}) and (\ref{[KV,conf] pm graviton}) agree with the $so(4,2)$ commutation relations. However, the commutator~(\ref{[conf,conf]pm graviton}) does not close on $so(4,2)$ as it is not equal to a Lie derivative with respect to the Killing vector $[V,V']$ (see Subsection~\ref{subsec_conf killing vectors}). The expression on the right-hand side of Eq.~(\ref{[conf,conf]pm graviton}) seems to describe a new symmetry transformation of $h_{\mu    \nu}$ that is second-order in derivatives. This symmetry transformation can be expressed more generally as
\begin{align}\label{higher symmetry pm graviton}
    D_{\xi} h_{\mu \nu} \equiv & ~\frac{3}{2} \left( \xi^{\rho}F_{\rho (\mu|\nu)} +\frac{1}{6}  \nabla_{\alpha}\xi_{\beta}\,\nabla_{(\mu} F^{\alpha \beta}_{\hspace{5mm}| \nu)}  
    \right) \\
    =& ~\xi^{\rho}F_{\rho (\mu|\nu)} +\frac{1}{4} \nabla_{(\mu}  \left( F^{\alpha   \beta}_{\hspace{4mm} | \nu)} \, \nabla_{\alpha} 
 \xi_{\beta}   \right),
\end{align}
where $\xi$ is any dS Killing vector, while in the second line, we have used~\cite{gravitation}
\begin{align}
\nabla_{\mu}\nabla_{\lambda}\xi_{\rho}=R_{\rho \lambda \mu\sigma} \xi^{\sigma}.
\end{align}

The algebra corresponding to the commutators~(\ref{[KV,KV]pm graviton})-(\ref{[conf,conf]pm graviton}) {would close on $so(4,2)$ if the new symmetry transformation $D_{\xi}h_{\mu \nu}$ [Eq.~(\ref{higher symmetry pm graviton})] could be expressed as $\mathcal{L}_{\xi}h_{\mu \nu}$ plus a gauge transformation or plus an expression that vanishes (off- or on-shell). To check this, we express $D_{\xi}h_{\mu \nu}$ as follows:
$$ D_{\xi} h_{\mu \nu} = \mathcal{L}_{\xi}h_{\mu   
 \nu} - \nabla_{(\mu}\xi^{\sigma} \, h_{\nu)  \sigma}+ \nabla_{(\mu}  \left( \frac{1}{4}   F^{\alpha   \beta}_{\hspace{4mm} | \nu)} \, \nabla_{\alpha} \xi_{\beta} - \xi^{\sigma}h_{\nu) \sigma}  \right),  $$
where the first term on the right-hand side is a Lie derivative. However, the last two terms do not seem to correspond to a gauge transformation (at least we were not able to express them in the form of a gauge transformation~(\ref{gauge transformations p.m. graviton})). Moreover, the last two terms do not vanish on-shell. This suggests that the symmetry algebra corresponding to the commutators (\ref{[KV,KV]pm graviton})-(\ref{[conf,conf]pm graviton}) does \textbf{not} close on $so(4,2)$. However, one could still wonder whether the non-closure on $so(4,2)$ is a gauge artifact and whether, when acting on the field strength, the second-order transformation $D_{\xi}$ reduces to a usual infinitesimal dS isometry transformation (i.e. a Lie derivative). However, after a straightforward calculation, we find that the field strength transforms as follows under the second-order transformation:
\begin{align}
    D_{\xi} F_{\alpha \beta | \gamma }  =  & \nabla_{\alpha} D_{\xi}h_{\beta   \gamma}   - \nabla_{\beta} D_{\xi}h_{\alpha   \gamma}  \\
     =& ~\mathcal{L}_{\xi} F_{\alpha \beta | \gamma } +\xi^{\rho} \left(  \frac{1}{4}  \nabla_{\rho}F_{\alpha   \beta | \gamma} - \nabla_{\gamma} F_{\alpha \beta | \rho}  \right) + \frac{1}{2}  \nabla_{\rho} \xi_{\lambda}~    g_{\gamma  [\alpha} ~F^{\rho \lambda}\,_{|\beta]} \nonumber \\
     & + \nabla_{[\alpha} \xi^{\rho}~ F_{\beta] \gamma | \rho} -\frac{5}{4} \nabla_{\gamma}\xi^{\rho} ~F_{\alpha \beta | \rho} +\frac{1}{4}  \nabla^{\sigma} \xi^{\rho}   ~\nabla_{\gamma}\nabla_{\sigma}F_{\alpha \beta | \rho}. \label{Dxi on field strength}
\end{align}
It is clear from this expression that $ D_{\xi} F_{\alpha \beta | \gamma }$ is not simply equal to a Lie derivative. Recalling that $D_{\xi}$ appears in the commutator between two unconventional conformal transformations as [see Eq. (\ref{[conf,conf]pm graviton})] 
$$ [\delta_{V} , \delta_{V'}] F_{\alpha \beta | \gamma} = D_{[V,V']}   F_{\alpha \beta | \gamma},$$
where $D_{[V ,V ']} F_{\alpha \beta | \gamma}$ is given by (\ref{Dxi on field strength}) with $\xi = [V ,V']$, indicates again that the algebra does \textbf{not} close on $so(4,2)$.}

Our results concerning the closure of the full symmetry algebra are inconclusive  - the study of the full algebra is something that we leave for future work. However, we note that the commutators $[D_{\xi} , D_{\xi'}]h_{\mu \nu}$ and $[D_{\xi} , \delta_{V}]h_{\mu \nu}$ seem to give rise to more new symmetry transformations that are higher-order in derivatives. Also, since $\mathcal{L}_{\xi}h_{\mu \nu}$ and $\delta_{V}h_{\mu \nu}$ are symmetries of the action, the commutator~(\ref{[conf,conf]pm graviton}) suggests that $D_{\xi} h_{\mu \nu}$ is also a symmetry of the action. 

{As $D_{\xi} h_{\mu \nu}$ is a higher derivative transformation, one might wonder whether this transformation, as well as the structure of the underlying symmetry algebra, is related to the higher symmetries of the Laplacian studied by Eastwood~\cite{Eastwood}. Unfortunately, it is not clear how the analysis of \cite{Eastwood} might be related to the results of our paper. In particular, in \cite{Eastwood} a symmetry of the Laplacian $\Delta$ is defined as a linear differential operator $\tilde{D}$ so that 
\begin{align} \label{def: Eastwood}
   \Delta \tilde{D} = \tilde{D}' \, \Delta 
\end{align}
for some linear differential operator $\tilde{D}'$. However, in our case where $\tilde{D}$ corresponds to, e.g., $\delta_{V}$ (\ref{unvonv conf pm graviton}) or $D_{\xi}$ (\ref{higher symmetry pm graviton}), the property (\ref{def: Eastwood}) (with $\tilde{D} = \tilde{D}'$) is satisfied only when acting on the on-shell partially massless graviton in the transverse-traceless gauge~(\ref{EOM max depth}), as 
$$\nabla^{\alpha} \nabla_{\alpha} \tilde{D}h_{\mu \nu} = \tilde{D}\nabla^{\alpha} \nabla_{\alpha} h_{\mu \nu}.$$
If the partially massless graviton is off-shell, then this property no longer holds. However, interestingly, a generalised version of (\ref{def: Eastwood}) is given by (\ref{transformation of Hmunu}), which holds off-shell, where the role of the Laplacian is played by the differential operator $H_{\mu \nu}$ appearing in the action (\ref{pm graviton action simple form}) of the partially massless graviton. To the best of our knowledge, (higher) symmetries of this type have not been classified and our paper seems to give the first discussion on them. To conclude, the partially massless graviton enjoys unconventional conformal symmetries (\ref{unvonv conf pm graviton}), while the underlying symmetry algebra does not close on $so(4,2)$ and its full structure is an open question.}

%%%%%%%%%%%%%%%%%%%%%%%%%%%%%%%%%%%%%%%%%%%%%%%%%%%%%%%%%%%%%%%%%%
\section{Unconventional conformal symmetry of the field equations~(\ref{EOM max depth}) for maximal depth partially massless spin-\texorpdfstring{$s \geq 2$}{s>2} fields} \label{sec_mac depth pm spin s>2}
In this Section, we will discuss the unconventional conformal symmetry of spin-$s\geq 2$ maximal depth partially massless fields at the level of the field equations~(\ref{EOM max depth}) only.

Let us define the rank-$(s+1)$ tensor $F_{\alpha \beta | \mu_{2} \mu_{3}...\mu_{s}} = F_{[\alpha \beta] | \mu_{2} \mu_{3}...\mu_{s}} =   F_{\alpha \beta | (\mu_{2} \mu_{3}...\mu_{s})}$ as
\begin{align}
    F_{\alpha \beta | \mu_{2} \mu_{3}...\mu_{s}} = ~2\,\nabla_{[\alpha} h_{\beta] \mu_{2}...\mu_{s}}.
\end{align}
We assume that $h_{\beta \mu_{2}...\mu_{s}}$ satisfies the field equations~(\ref{EOM max depth}) and it immediately follows that $$g^{\alpha \mu_{2}   } F_{\alpha \beta | \mu_{2} \mu_{3}...\mu_{s}} = g^{\alpha \mu_{3}   } F_{\alpha \beta | \mu_{2} \mu_{3}...\mu_{s}}= ...= g^{\alpha \mu_{s}   } F_{\alpha \beta | \mu_{2} \mu_{3}...\mu_{s}} =0,$$
while, using Eq.~(\ref{commut of cov derivs}), it is easy to show that
\begin{align}\label{div freedom spin s>2 field strength}
   \nabla^{\alpha  } F_{\alpha \beta | \mu_{2} \mu_{3}...\mu_{s}} =0 .
\end{align}
$F_{\alpha \beta | \mu_{2}...\mu_{s}}$ is also traceless with respect to any pair of $\mu_{i},\mu_{j}$ indices ($i, j \in \{ 2,3,...,s  \}$).
The following identities are satisfied off-shell:
\begin{align}
  &  F_{\alpha \beta | \mu_{2} \mu_{3}...\mu_{s}} +    F_{\mu_{2}\alpha | \beta \mu_{3}...\mu_{s}} +   F_{\beta  \mu_{2} | \alpha\mu_{3}...\mu_{s}} =0, \label{alg Bianchi}\\
  & \nabla_{[\kappa}     F_{\alpha \beta] | \mu_{2} \mu_{3}...\mu_{s}} = 0 \label{diff Bianchi}.
\end{align}
From Eqs.~(\ref{div freedom spin s>2 field strength}) and (\ref{alg Bianchi}) it also follows that $F_{\alpha \beta | \mu_{2}...\mu_{s}}$ is divergence-free with respect to any index.
%%%%%%%%%%%%%%%%%%%%%%%%%%%%%%%%%%%%%%%%%%%%%%%%%%%%%%%%%%%%%%%%%%%%%%%%%%%%%%%%%%%%%%%%%%%%%%%%%%%%%%%%%%%%%%%%%%%%%%%%
\subsection{Unconventional conformal transformations for spin \texorpdfstring{$s \geq 2$}{s>=2}} \label{subsec_unconf conf spin s>=2}
The unconventional infinitesimal conformal transformations of the maximal depth partially massless fields of spin-$s \geq 2$ are given by
\begin{align} \label{unvonv conf pm spin-s>2}
    \delta_{V} h_{\mu_{1}...\mu_{s} } =V^{\rho} F_{\rho (\mu_{1} | \mu_{2}...\mu_{s})} = V^{\rho}  \left( \nabla_{\rho}h_{\mu_{1} ...\mu_{s}}- \nabla_{(\mu_{1}}h_{\mu_{2}...\mu_{s})  \rho}     \right),
\end{align}
where $V^{\rho}$ is any genuine conformal Killing vector~(\ref{CKV equation}) of $dS_{4}$. Let us briefly prove that if $h_{\mu_{1} ... \mu_{s}}$ satisfies the field equations~(\ref{EOM max depth}), then so does $\delta_{V} h_{\mu_{1} ... \mu_{s}}$. Using the properties of $F_{\rho \mu_{1} | \mu_{2}...\mu_{s}}$ it is easy to show that  $\delta_{V} h_{\mu_{1}...\mu_{s} }$ is traceless with respect to any pair of indices. Also, using Eq.~(\ref{properties of phi 1}) and the properties of $F_{\rho \mu_{1} | \mu_{2}...\mu_{s}}$, we immediately find $\nabla^{\mu_{1}}\delta_{V} h_{\mu_{1}...\mu_{s} } = 0$. Finally, a straightforward calculation gives
$$ \nabla^{\kappa}  \nabla_{\kappa}  \delta_{V} h_{\mu_{1}...\mu_{s} }  = V_{\rho} \left( - 1 + \nabla_{\kappa}\nabla^{\kappa}  \right) F^{\rho}_{\hspace{2mm} (\mu_{1} | \mu_{2}...\mu_{s})}.$$ In order to simplify this expression we can use
\begin{align*}
 \nabla_{\kappa}\nabla^{\kappa}  F^{\rho}_{\hspace{2mm}(\mu_{1} | \mu_{2}...\mu_{s})} =& \nabla_{\kappa}\left(-\nabla_{(\mu_{1}}  F^{\kappa \rho}_{\hspace{4mm} | \mu_{2}...\mu_{s})} + \nabla^{\rho}F^{\kappa}_{\hspace{2mm}(\mu_{1} | \mu_{2}...\mu_{s})} \right) \\
 &= (s+3)\, F^{\rho} \,_{(\mu_{1} | \mu_{2}...\mu_{s})},   
\end{align*}
where in the first line we used Eq.~(\ref{diff Bianchi}), while in the second line we commuted the covariant derivatives and we used (\ref{div freedom spin s>2 field strength}). Thus, we arrive at 
$$ \nabla_{\kappa}  \nabla^{\kappa}  \delta_{V} h_{\mu_{1}...\mu_{s} }  =(s+2)\, \delta_{V} h_{\mu_{1}...\mu_{s} }.$$
We have thus verified that $\delta_{V}h_{\mu_{1} ... \mu_{s}}$ is a symmetry of the field equations~(\ref{EOM max depth}) for maximal depth partially massless field of spin $s \geq 2$.
%%%%%%%%%%%%%%%%%%%%%%%%%%%%%%%%%%%%%%%%%%%%%%%%%%%%%%%%%%%%%%%%%%%%%
\subsection{The algebra does not close on \texorpdfstring{$so(4,2)$}{so(4,2)}}\label{subsec_algebra pm max depth spin s>2}
In order to understand the structure of the symmetry algebra we work as in the case of the partially massless graviton in Subsection~\ref{subsec_algebra pm graviton}. We find
\begin{align}
    [\mathcal{L}_{\xi},  \mathcal{L}_{\xi '} ] h_{\mu_{1}...\mu_{s}} &= \mathcal{L}_{ [\xi, \xi']}h_{\mu_{1} ... \mu_{s}}, \label{[KV,KV]pm spin s>2} \\
    [\mathcal{L}_{\xi},  \delta_{V} ] h_{\mu_{1} ... \mu_{s}}=  & ~ \mathcal{L}_{\xi} \, \delta_{V}h_{\mu_{1} ... \mu_{s}}- \delta_{V}  \mathcal{L}_{\xi} h_{\mu_{1}...\mu_{s}}\nonumber   \label{[KV,conf] pm spin s>2}  \\
    =& ~\delta_{[\xi ,V ]}h_{\mu_{1} .. \mu_{s}}, \\
    [\delta_{V},  \delta_{V'} ] h_{\mu_{1} ... \mu_{s}}=  & ~ \delta_{V} \, \delta_{V'}h_{\mu_{1} ... \mu_{s}}- \delta_{V'} \, \delta_{V}h_{\mu_{1} ... \mu_{s}}\nonumber\\
    =& ~ D_{[V,V']} h_{\mu_{1} ... \mu_{s}}\label{[conf,conf] pm spin s>2}.
\end{align}
The symmetry transformation on the right-hand side of Eq.~(\ref{[conf,conf] pm spin s>2}) is second-order in derivatives (i.e., it is the higher-spin generalisation of (\ref{higher symmetry pm graviton})). It is given by the following equivalent expressions:
\begin{align}\label{higher symmetry pm spin s>2}
    &D_{\xi} h_{\mu_{1} ... \mu_{s}} \equiv  ~\frac{s+1}{s} \left( \xi^{\rho}F_{\rho (\mu_{1}|\mu_{2}...\mu_{s})} +\frac{1}{2(s+1)}  \nabla_{\alpha}\xi_{\beta}\,\nabla_{(\mu_{1}} F^{\alpha \beta}_{\hspace{5mm}| \mu_{2}...\mu_{s})}  
    \right) \\
    =& ~\xi^{\rho}F_{\rho (\mu_{1}|\mu_{2}...\mu_{s})} +\frac{1}{2s} \nabla_{(\mu_{1}}  \left( F^{\alpha   \beta}_{\hspace{4mm} | \mu_{2}...\mu_{s})} \, \nabla_{\alpha} 
 \xi_{\beta}   \right)\\
 =&  \mathcal{L}_{\xi}h_{\mu_{1}...   
 \mu_{s}} -(s-1) \nabla_{(\mu_{1}}\xi^{\sigma} \, h_{\mu_{2}...\mu_{s})  \sigma}+ \nabla_{(\mu_{1}}  \left( \frac{1}{2s}   F^{\alpha   \beta}_{\hspace{4mm} | \mu_{2}...\mu_{s})} \, \nabla_{\alpha} \xi_{\beta} - \xi^{\sigma}h_{\mu_{2}...\mu_{s}) \sigma}  \right).
\end{align}
For $s=1$ (i.e., in the Maxwell case), $D_{\xi}h_{\mu_{1}}$ reduces to a Lie derivative plus a gauge transformation, reflecting the well-known $so(4,2)$ symmetry of the Maxwell theory discussed in the Introduction. From the commutation relations~(\ref{[KV,KV]pm spin s>2})-(\ref{[conf,conf] pm spin s>2}) [and from the form of $D_{\xi}h_{\mu_{1}...\mu_{s}}$ in Eq.~(\ref{higher symmetry pm spin s>2})], we conclude that the symmetry algebra for the field equations~(\ref{EOM max depth}) does \textbf{not} close on $so(4,2)$ (the arguments are the same as in the case of the partially massless graviton discussed in Subsection~\ref{subsec_algebra pm graviton}).

As in the case of the partially massless graviton, our results concerning the closure of the symmetry algebra are inconclusive. However, we again note that the commutators $[D_{\xi} , D_{\xi'}]h_{\mu_{1} ... \mu_{s}}$ and $[D_{\xi} , \delta_{V}]h_{\mu_{1}   .... \mu_{s}}$ seem to give rise to more new symmetry transformations that are higher-order in derivatives.
%%%%%%%%%%%%%%%%%%%%%%%%%%%%%%%%%%%%%%%%%%%%%%%%%%%%%%%%%%%%%%%%%%%%%%%%%%%%%%%%%%%%%%%%%%%%%%%%%%%%%%%%%%%%%%%
\section{Supermultiplet of complex partially massless spin-2 field and complex spin-3/2 field and unconventional conformal symmetry} \label{sec_susy}
(Our notation and conventions for fermionic fields and gamma matrices on $dS_{4}$ are as in Refs.~\cite{Letsios_Hidden, Letsios_announce}.)

\noindent The only known examples of unbroken unitary supersymmetric field theories on $dS_{4}$ correspond to the superconformal field theories studied in Ref.~\cite{dS revisited} - see also appendix A of Ref.~\cite{Hristov}. (For discussions on $dS_{2}$ supergravity see Refs.~\cite{Beatrix, Damian}.) The main obstacles for SUSY in $dS_{4}$ are the absence of reality conditions for the standard dS Killing spinors, and the non-existence of unitary representations of the usual dS superalgebra~\cite{Nieuwenhuizen, Lukierski}. In Ref.~\cite{dS revisited}, these obstacles were avoided because conformal Killing spinors of $dS_{4}$ were used (conformal Killing spinors admit reality conditions), while also the commutator of two SUSY variations closed on $so(4,2)$ instead of $so(4,1)$.

In this Section, we will uncover a new supermultiplet containing two free complex (classical) fields on $dS_{4}$ that will showcase the role of the unconventional conformal symmetries, as well as the role of the corresponding algebra discussed in Section~\ref{sec_pm graviton}. In particular, we will present the SUSY transformations for the supermultiplet of a complex partially massless spin-2 field $\mathcal{A}_{\mu \nu} = \mathcal{A}_{(\mu \nu)}$ and a Dirac vector-spinor field $\psi_{\mu}$. We will focus on demonstrating the realisation of a SUSY representation on the solution space of the classical field equations. We leave a fuller and quantum field theoretic study of this theory for future work.

As for dealing with the main obstacles for SUSY in $dS_{4}$, we `ignore' the `reality condition obstacle' since we study a complex supermultiplet. This means that we will use the standard complex dS Killing spinors $\epsilon$. These satisfy~\cite{dS revisited}
\begin{align}\label{dS Killing spinors}
  \left( \nabla_{\mu}+\frac{i}{2} \gamma_{\mu}  \right) \epsilon = 0 .
\end{align}
(As we will see below, although the Killing spinors $\epsilon$ are complex, only the real parts of Killing spinor bilinears appear in the commutators of two SUSY variations!)
Also, although the unitarity of our theory is not investigated here, the fact that the commutator of two SUSY variations closes on an algebra that is larger than $so(4,1)$ ensures that the non-unitarity related to dS superalgebras~\cite{Nieuwenhuizen, Lukierski} is avoided. However, although it is clear that there is no closure on $so(4,1)$ or $so(4,2)$, we note that our results are inconclusive concerning the closure of the full even subalgebra of our complex supermultiplet. This inconclusiveness occurs because the even subalgebra commutation relations - or, to be precise, the part of the commutation relations that we have investigated - coincide with the commutators~(\ref{[KV,KV]pm graviton}) - (\ref{[conf,conf]pm graviton}) (but with $h_{\mu \nu}$ replaced by $\mathcal{A}_{ \mu \nu}$) and their study has yet to be completed.
%%%%%%%%%%%%%%%%%%%%%%%%%%%%%%%%%%%%%%%%%%%%%%%%%%%%%%%%%%%%%%%%%%%%%%%%%%%%%%%%%%%%%%%%%%%%%%%%%%%%%%%%%%%%%%%%%%%%%%%
 $$ \textbf{Supermultiplet of complex partially massless spin-2 field and spin-3/2 field}   $$

\noindent Consider a complex partially massless spin-2 field $\mathcal{A}_{\mu \nu}$ satisfying Eqs.~(\ref{EOM max depth}):
\begin{align} \label{EOM pm complex graviton}
    &\Big( \nabla^{\alpha} \nabla_{\alpha}-4 \Big)\mathcal{A}_{\mu \nu} = 0, \nonumber\\
    & \nabla^{\alpha} \mathcal{A}_{\alpha \mu}=0, \hspace{4mm} g^{\alpha \beta} \mathcal{A}_{\alpha \beta}=0.
\end{align}
The field $\mathcal{A}_{\mu \nu}$ has 4 (complex) propagating degrees of freedom.
For later convenience, note that Eqs.~(\ref{EOM pm complex graviton}) enjoy the symmetries of the real field equations~(\ref{EOM max depth}).
In particular, Eqs.~(\ref{EOM pm complex graviton}) are invariant under on-shell gauge transformations of the form
\begin{align*}
     \delta \mathcal{A}_{\mu \nu} = \left( \nabla_{(\mu}  \nabla_{\nu)} +g_{\mu   \nu}   \right)b, 
\end{align*}
where $b$ is a complex scalar gauge function satisfying
\begin{align*}
     &\Box \,b = -4\,b.
 \end{align*}
Also, they are invariant under infinitesimal dS transformations, $\mathcal{L}_{\xi}\mathcal{A}_{\mu \nu}$ [Eq.~(\ref{Lie derivative})], and unconventional conformal transformations~(\ref{unvonv conf pm graviton}),
 \begin{align} \label{unvonv conf complex graviton}
     \delta_{V} \mathcal{A}_{\mu \nu} = V^{\rho} \mathcal{F}_{\rho (\mu | \nu)}   ,
 \end{align}
where we have denoted the complex field strength as
\begin{align}
    \mathcal{F}_{\mu \nu | \alpha} = \nabla_{\mu} \mathcal{A}_{\nu \alpha }   - \nabla_{\nu} \mathcal{A}_{ \mu \alpha }.
\end{align}
The commutators of the dS and unconventional conformal transformations are given by Eqs.~(\ref{[KV,KV]pm graviton})-(\ref{[conf,conf]pm graviton}) with $h_{\mu \nu}$ replaced by $\mathcal{A}_{\mu \nu}$ (and ${F}_{\mu \nu | \alpha}$ replaced by $\mathcal{F}_{\mu \nu | \alpha}$). Let us also write down the second-order transformation~(\ref{higher symmetry pm graviton}) for the complex partially massless spin-2 field:
\begin{align}\label{higher symmetry pm graviton complex}
    D_{\xi} \mathcal{A}_{\mu \nu} = & ~\frac{3}{2} \left( \xi^{\rho}\mathcal{F}_{\rho (\mu|\nu)} +\frac{1}{6}  \nabla_{\alpha}\xi_{\beta}\,\nabla_{(\mu} \mathcal{F}^{\alpha \beta}_{\hspace{5mm}| \nu)}     \right).
\end{align}

Consider also the complex vector-spinor $\psi_{\mu}$ satisfying
\begin{align}\label{EOM spin-3/2}
  & \slashed{\nabla} \psi_{\mu} = 0  \nonumber \\
 & \gamma^{\mu} \psi_{\mu} =   \nabla^{\mu}  \psi_{\mu} = 0.
\end{align}
Note that the field $\psi_{\mu}$ is not a fermionic gauge potential, i.e. it is not a strictly massless spin-3/2 field (gravitino), as the latter has an imaginary mass parameter~\cite{Letsios_announce}. The number of (complex) propagating degrees of freedom for $\psi_{\mu}$ is 4~\cite{Deser_Waldron_phases}.

Since each of the fields $\mathcal{A}_{\mu \nu}$ and $\psi_{\mu}$ has 4 (complex) degrees freedom, it is tempting to look for SUSY transformations, $\delta^{susy}_{\epsilon} \mathcal{A}_{\mu \nu}$ and $\delta^{susy}_{\epsilon} \mathcal{\psi}_{\mu}$, that transform the classical solution spaces of Eqs.~(\ref{EOM pm complex graviton}) and (\ref{EOM spin-3/2}) into each other. We have found that the desired SUSY transformations are: 
\begin{align}
  &  \delta^{susy}_{\epsilon}\mathcal{A}_{\mu \nu} = \overline{\epsilon} \gamma_{( \mu} \psi_{\nu)} +\frac{2i}{5} \overline{\epsilon} \nabla_{( \mu} \psi_{\nu)} , \label{susy delta Amn} \\
  & \delta^{susy}_{\epsilon} \psi_{\mu} = \frac{1}{4}  \nabla_{\lambda}\mathcal{A}_{\mu \sigma}   \gamma^{\sigma  \lambda}  \epsilon = \frac{1}{8}  \mathcal{F}_{ \lambda \sigma | \mu}  \, \gamma^{\sigma \lambda}  \epsilon, \label{susy delta psim}
\end{align}
where $\gamma^{\sigma  \lambda} = \gamma^{[\sigma} 
  \gamma^{\lambda]}$~\cite{Freedman}, while $\epsilon$ is a complex Killing spinor~(\ref{dS Killing spinors}). In particular, a straightforward calculation shows that if $\psi_{\mu}$ satisfies Eqs.~(\ref{EOM spin-3/2}), then $ \delta^{susy}_{\epsilon}  \mathcal{A}_{\mu \nu}$ satisfies Eqs.~(\ref{EOM pm complex graviton}). Similarly, if $\mathcal{A}_{\mu \nu}$ satisfies Eqs.~(\ref{EOM pm complex graviton}), then $\delta^{susy}_{\epsilon}  \psi_{\mu }$ satisfies Eqs.~(\ref{EOM spin-3/2}). This demonstrates that the fields $\mathcal{A}_{\mu \nu}$ and $\psi_{\mu}$ form a supermultiplet (i.e. we have a representation of SUSY on the solution space of the field equations - see, e.g.~\cite{Freedman, West}).

The role of the unconventional conformal transformations~(\ref{unvonv conf complex graviton}) in our complex supermultiplet manifests itself by computing the commutator of two SUSY variations $[\delta^{susy}_{\epsilon_{2}} , \delta^{susy}_{\epsilon_{1}}]$. After a straightforward calculation, we find
\begin{align}\label{[susy, susy]pm grav compl}
  \frac{5}{2}  [\delta^{susy}_{\epsilon_{2}}, \delta^{susy}_{\epsilon_{1}}   ]\mathcal{A}_{\mu \nu}=  \delta_{V'}\mathcal{A}_{\mu \nu}-i \frac{3}{2} \left( \xi^{' \rho}~^{*}\mathcal{F}_{\rho (\mu|\nu)} +\frac{1}{6}  \nabla_{\alpha}\xi'_{\beta}\,\nabla_{(\mu} \,^{*}\mathcal{F}^{\alpha \beta}_{\hspace{5mm}| \nu)}     \right) .
\end{align}
Interestingly, the first term on the right-hand side ($\delta_{V'}\mathcal{A}_{\mu \nu}$) is an unconventional conformal transformation~(\ref{unvonv conf complex graviton}) with respect to the real genuine conformal Killing vector
\begin{align}
    V^{'\rho} \equiv \frac{1}{2}  \left( \overline{\epsilon_{1}}  \gamma^{\rho} \epsilon_{2}    - \overline{\epsilon_{2}}  \gamma^{\rho} \epsilon_{1}\right) = \frac{1}{2}  \left( \overline{\epsilon_{1}}  \gamma^{\rho} \epsilon_{2}    + ( \overline{\epsilon_{1}}  \gamma^{\rho} \epsilon_{2})^{\dagger}\right).
\end{align}
The second term on the right-hand side of Eq.~(\ref{[susy, susy]pm grav compl}) corresponds to the second-order symmetry $D_{\xi'}$~(\ref{higher symmetry pm graviton complex}), but with the field strength replaced by its dual $$^{*}\mathcal{F}_{\rho \mu|\nu} = \frac{1}{2}  \epsilon_{\rho \mu}^{\hspace{4mm}\alpha \beta}\mathcal{F}_{\alpha \beta |\nu}.$$
The real Killing vector $\xi'$ is given by
\begin{align}
    \xi^{'\rho} \equiv \frac{1}{2}  \left( \overline{\epsilon_{1}} \gamma^{5} \gamma^{\rho} \epsilon_{2}    - \overline{\epsilon_{2}} \gamma^{5} \gamma^{\rho} \epsilon_{1}\right) = \frac{1}{2}  \left( \overline{\epsilon_{1}} \gamma^{5} \gamma^{\rho} \epsilon_{2}    + ( \overline{\epsilon_{1}}  \gamma^{5}\gamma^{\rho} \epsilon_{2})^{\dagger}\right).
\end{align}

The commutator of two SUSY variations~(\ref{[susy, susy]pm grav compl}) does not close on $so(4,2)$, while the full structure of the algebra is currently unknown [see Eqs.~(\ref{[KV,KV]pm graviton})-(\ref{[conf,conf]pm graviton}) and the discussion in the passage below them]. We leave the computation of the commutator $[\delta^{susy}_{\epsilon_{2}}, \delta^{susy}_{\epsilon_{1}}] \psi_{\mu}$ and the study of the full algebra for future work. Note that, once one determines $[\delta^{susy}_{\epsilon_{2}}, \delta^{susy}_{\epsilon_{1}}] \psi_{\mu}$, it is natural to find the analogue of the unconventional conformal transformations for the spin-3/2 field.

{Before concluding, we note that non-unitary supermultiplets containing partially massless fields on $AdS_{4}$ have been studied in Refs.~\cite{Hinterbichler, Hinterbichler2}. The nature of our superalgebra is different from the nature of the superalgebras in Refs.~\cite{Hinterbichler, Hinterbichler2}, as in these references the commutators of two SUSY variations close on AdS isometries, while in our case the corresponding commutators do \textbf{not} close on dS isometries (moreover, unlike the theories presented in~\cite{Hinterbichler, Hinterbichler2}, our supermultiplet is likely to be unitary, as discussed earlier in this Section).\footnote{Off-shell partially massless supermultiplets in four-dimensional $\mathcal{N}=1$ AdS superspace were presented for the first time in Ref.~\cite{Hutchings}.} For discussions on superconformal theories in de Sitter see also~\cite{AdS/dS}, while the imprints of higher spin supermultiplets on cosmological correlators have been discussed in~\cite{SUSY CMB}}.

Open questions that arise from our findings are discussed in the next Section.
%%%%%%%%%%%%%%%%%%%%%%%%%%%%%%%%%%%%%%%%%%%%%%%%%%%%%%%%%%%%%%%%%%%%%%%%%%%%%%%%%%%%%%%5

\section{Discussion and Open Questions}\label{sec_discussion}
Let us summarise our findings and present the open questions arising from them.
\begin{itemize}
\item We presented the unconventional conformal symmetry of the partially massless graviton both at the level of the action and at the level of the field equations in the transverse-traceless gauge~(see Section~\ref{sec_pm graviton}). In the case of partially massless spin-$s > 2$ fields of maximal depth, this was achieved only at the level of the equations of motion in the transverse-traceless gauge. Whether the covariant actions for these fields are invariant under the unconventional conformal symmetries~(\ref{unvonv conf pm spin-s>2}) is still an open question. Also, the structure of the symmetry algebra for all spin-$s \geq 2$ maximal depth partially massless fields is currently unknown, as our results could not determine whether the algebra closes (see Subsections~\ref{subsec_algebra pm graviton} and \ref{subsec_algebra pm max depth spin s>2}). However, our findings suggest that partially massless bosons of maximal depth do not enjoy $so(4,2)$ symmetry in agreement with \cite{Conf_Bekaert}. Moreover, expressions for the Noether currents associated with the unconventional conformal symmetries are missing from the literature.

    \item In Section~\ref{sec_susy}, a supermultiplet containing a spin-2 partially massless field on $dS_{4}$ was presented for the first time (where the commutator of two SUSY variations does not close either on $so(4,1)$ or $so(4,2)$). The analysis was focused on the realisation of SUSY at the level of the field equations. It would be interesting to extend this at the level of the Lagrangian. Also, the quantisation of the theory, as well as the study of its full algebra and unitarity, are interesting open problems. However, as mentioned earlier, the fact that the commutator of two SUSY transformations (\ref{[susy, susy]pm grav compl}) does not close on $so(4,1)$ suggests that the theory is likely to be unitary.

    \item The supermultiplet discussed in Section~\ref{sec_susy} is complex. Can we find a real version of this theory using Majorana conformal Killing spinors (as in Ref.~\cite{dS revisited}) instead of standard complex Killing spinors~(\ref{dS Killing spinors})? Also, it is worth wondering whether we can introduce interactions while maintaining SUSY. Note that the first consistent (non-supersymmetric) interacting theory of a real partially massless spin-2 field and a massive (Majorana) spin-3/2 field on $dS_{4}$ was recently constructed in Ref.~\cite{Boulanger} (see also Ref.~\cite{Zinoviev}).

\item It is reasonable to expect that the complex supermultiplet of Section~\ref{sec_susy} can be extended to higher spins. In particular, based on the counting of the degrees of freedom, it is possible that a complex supermultiplet containing a bosonic spin-$s >2$ maximal depth partially massless field and a spin-$(s-1/2) > 3/2$ tensor-spinor exists. We leave the investigation of this question for future work. 

\end{itemize}
\acknowledgments
The author thanks Atsushi Higuchi, Dionysios Anninos, F.F. John, Alan Rios Fukelman, Jeremy Mann, Simon Ekhammar, Benoit Vicedo, Guillermo `il professore' Silva, and Mati `il alumni' Sempe for useful discussions. The author is deeply grateful to Eleni Gagon and Beth E. Le Friant, as well as to the Eleni Gagon Survivor's Trust, for supporting the present work. 

\noindent

%%%%%%%%%%%%%%%%%%%%%%%%%%%%%%%%%%%%%%%%%%%%%%%%%%%%%%%%%%%%%%%%%5%%%%%%%%
\appendix

%%%%%%%%%%%%%%%%%%%%%%%%%%%%%%%%%%%%%%%%%%%%%%%

\nocite{*}

%%%%%%%%%%%%%%%%%%%%%%%%%%%%%%%%%%%%%%%%%%%%%%%%%%%%%%%%%%%%%%%%%%%%%%%%%%%%%%%%%%%%%%%%%%%%%%%%%%%%%%%%%%%%%%%%%%%%%%%%%%%%%%%%%%%%%%%%%%%%%%%5555555555555555555555555555%%%
\providecommand{\noopsort}[1]{}\providecommand{\singleletter}[1]{#1}%

\end{document}